\documentclass[sigconf,nonacm]{acmart}

\setcopyright{none}
\renewcommand\footnotetextcopyrightpermission[1]{}
\AtBeginDocument{%
  }

\usepackage{amsmath}
\usepackage{booktabs}
\usepackage{bm}
\usepackage{array}
\usepackage{graphicx}
\usepackage{placeins}
\usepackage{algorithm}
\usepackage{algpseudocode}
\title{Beyond a Scalar: Distributional Serving Interfaces for Watch-Time Prediction}

\author{Xuan Liu}
\authornote{Code available at \url{https://github.com/Xuanxuana1/DSI}.}
\affiliation{%
  \institution{Shanghai Jiao Tong University}
  \city{Shanghai}
  \country{China}}
\email{liuxuan_cn@outlook.com}

\author{Jingbin Qian}
\affiliation{%
  \institution{Rice University}
  \city{Houston}
  \state{Texas}
  \country{USA}}
\email{jingbinqian2002@gmail.com}

\author{Zhanyu Liu}
\affiliation{%
  \institution{Shanghai Jiao Tong University}
  \city{Shanghai}
  \country{China}}
\email{zhyliu00@sjtu.edu.cn}

\author{Hefeng Zhou}
\affiliation{%
  \institution{Shanghai Jiao Tong University}
  \city{Shanghai}
  \country{China}}
\email{h.rezin.zhou@gmail.com}

\begin{document}

\begin{abstract}
Watch time is the primary engagement signal in short video feeds, and its prediction directly affects ranking and exposure.
Existing methods improve watch time prediction by correcting duration bias or modeling richer distributions, but most expose only an expected or debiased watch time at serving time.
Even when video duration is available to later models, the interface gives only one estimate of watch time and no probabilities for completion, overplay, or other regions relevant to downstream tasks.
To address this limitation, we propose the Distributional Serving Interface (DSI), which has a distribution provider, a compact, low-dimensional summary, and lightweight readouts tailored to each task.
The provider learns a joint distribution over four watch states derived from watch ratio and their event times; rules based on video duration remove incompatible combinations, while a restoration loss preserves accuracy in seconds.
The summary reduces this distribution to a small set of event probabilities, time scales relative to duration, and uncertainty statistics.
After training the provider, we fix its parameters and train value and ranking readouts that combine the summary with raw context.
Across KuaiRec, KuaiRand-1K, and WeChat21, the complete DSI system achieves the lowest MAE on all three datasets, beating the strongest result among nine baselines by $1.9\%$ to $8.5\%$, and achieves the best XAUC on two.
It also leads retrieval metrics that account for video duration when complete systems are compared.
With matched readouts held constant, the summary retains information relevant to each task beyond a predicted mean paired with video duration.
Using the same lightweight linear heads for each new target, it also performs best on two new watch-time targets and improves a separately logged engagement target, while a randomly initialized provider does not reproduce this gain.
\end{abstract}



\maketitle


\section{Introduction}
\label{sec:intro}

Short-video platforms have become one of the dominant forms of online media consumption, serving billions of video plays per day~\citep{zhan2022d2q,deng2025onerec}.
Their feeds autoplay: a video starts playing without a click, so click-through signals carry little information about user interest~\citep{sun2024cread}.
Watch time has therefore become the primary engagement feedback on these platforms~\citep{covington2016youtube,yi2014beyond}: it is continuous, logged for nearly every impression, and its prediction directly shapes ranking and exposure allocation~\citep{yang2026dadf}.

\begin{figure}[t]
\centering
\includegraphics[width=0.70\columnwidth]{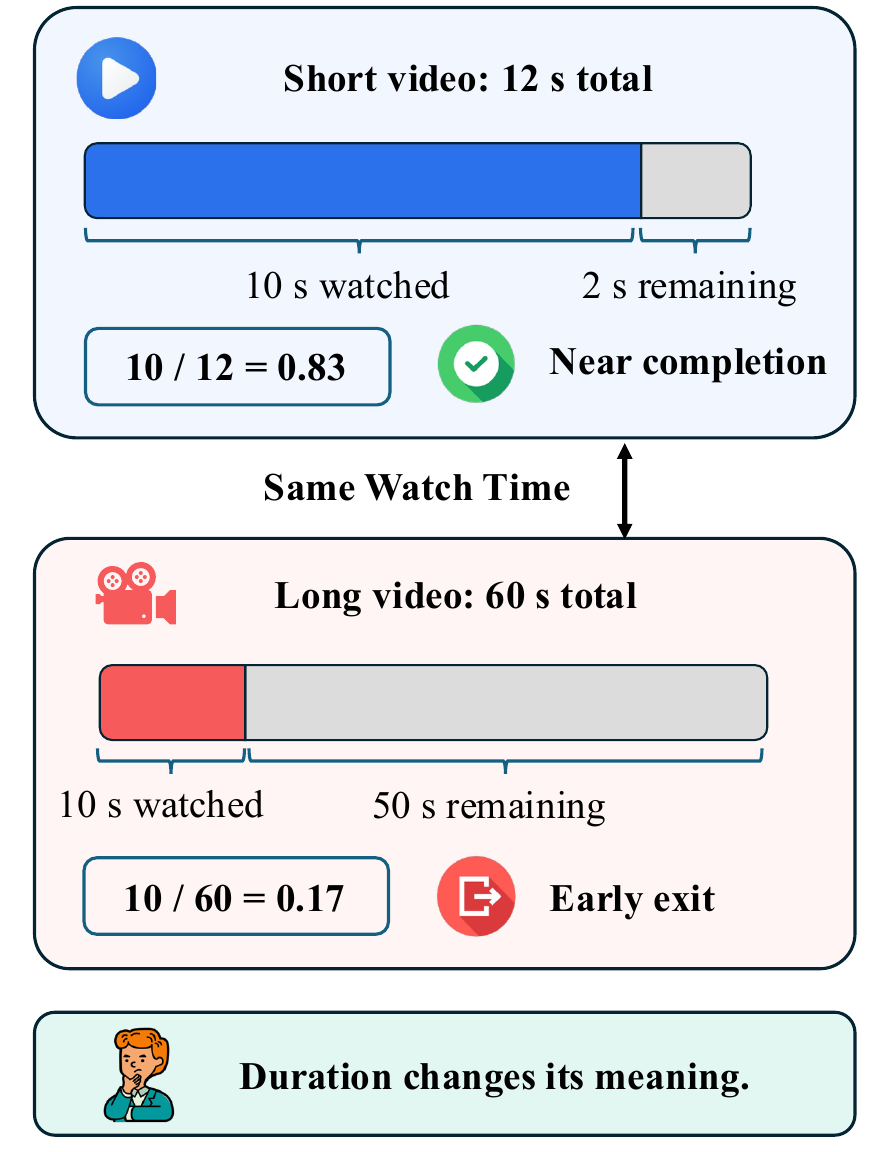}
\caption{The same 10 watched seconds carries different duration-relative meanings: near completion for a 12-second video, but an early exit from a 60-second video.}
\Description{Two stacked duration timelines showing that ten seconds is near completion for a short video but an early exit for a long video.}
\label{fig:duration-contrast}
\vspace{-0.8em}
\end{figure}

Existing predictors refine how this signal is fitted, correcting duration bias in the logged value~\citep{zhan2022d2q,zhao2024cwm} or modeling a richer watch-time distribution~\citep{lin2023tpm,egmn2025}, and several of them maintain a rich distribution internally during training.
At serving time, however, their estimator output is usually one expected or debiased watch time per impression, which downstream decisions combine with available context such as video duration.
Ten seconds indicates near completion for a 12-second video but an early exit from a 60-second one (Figure~\ref{fig:duration-contrast}).
A downstream layer with access to duration can resolve this immediate ambiguity by normalizing the point prediction.
Yet the resulting pair of predicted watch time and duration still exposes only one conditional mean.
Different conditional distributions can share that mean and duration while assigning different probability mass to completion, overplay, and other task-relevant regions.
This limitation becomes more restrictive after deployment, when new decisions must either reuse the same point estimate or retrain the estimator against another objective.
Existing methods improve watch-time estimation, but they do not provide a reusable serving output that retains \emph{what kind} of outcome may occur and \emph{when} it may occur beyond a duration-aware conditional mean.

To address this limitation, we propose the distributional serving interface (DSI), a three-stage framework that turns an estimated distribution of watch outcomes into a reusable serving output.
The first stage derives four operational event types and their time targets from logged watch time and video duration, providing distributional supervision without requiring a new feedback label.
The second stage trains a provider to estimate the joint distribution over event type and event time.
Support constraints based on video duration exclude incompatible outcomes, while a watch-time restoration loss preserves prediction accuracy in seconds.
The third stage freezes the trained provider and converts its distribution into a compact, low-dimensional summary of event probabilities, absolute and relative time scales, and uncertainty.
Lightweight value and ranking readouts combine this summary with raw context, and a task introduced later can learn a new readout without retraining the provider.

The contributions of this work are summarized as follows.
\begin{itemize}
\item We study watch-time prediction as a serving-interface problem. Even when video duration is available downstream, a scalar prediction does not expose the distributional information that different current or future decisions may require.
\item We propose DSI, which freezes a distribution provider and exposes a compact, low-dimensional summary to lightweight task-specific readouts. We evaluate it with a three-stage typed instantiation that derives operational watch outcomes and models their event times.
\item Experiments on three public datasets show that the complete DSI system achieves the lowest MAE among nine baselines on all three datasets and the best XAUC on two. Comparisons with matched readouts and reuse experiments further assess what the distributional interface contributes beyond a duration-aware scalar prediction.
\end{itemize}


\section{Related Work}
\label{sec:related}

\paragraph{Watch-time prediction.}
Watch time is a dense implicit-feedback signal for video recommendation~\citep{covington2016youtube,yi2014beyond}.
One line corrects duration bias in the logged value, through duration-conditioned quantiles~\citep{zhan2022d2q}, watch-time gain~\citep{zheng2022dvr}, denoising~\citep{zhao2023d2co}, counterfactual watch time~\citep{zhao2024cwm}, or duration-invariant features~\citep{jin2026difl}.
A second line models the watch-time distribution, through ordinal trees~\citep{lin2023tpm}, adaptive bins~\citep{sun2024cread}, conditional quantiles~\citep{lin2024cqe}, autoregressive tokens~\citep{ma2026gr}, or exponential--Gaussian mixtures~\citep{egmn2025}.
These methods differ in parameterization and supervision, and several retain a rich internal watch-time distribution.
Discrete competing-risks models also estimate a joint distribution over event type and event time~\citep{lee2018deephit}; DSI instead uses deterministic operational partitions of watch time rather than independently observed event causes.
DSI differs in the object exposed downstream: rather than stopping at a duration-aware point estimate, it keeps a compact summary of the estimated distribution available to task-specific readouts.
Relative-advantage debiasing similarly separates distribution estimation from preference learning~\citep{liu2025debiasing}; the distinction is discussed in the Remark following Section~\ref{sec:readout}.
The duration-dependent support of our typed likelihood follows coarse-data and interval-censoring theory~\citep{heitjan1991coarse,turnbull1976edf,zhang2010interval}.

\paragraph{From prediction to decision.}
Collapsing a rich representation to one fixed quantity is one instance of a familiar gap: minimizing prediction error and choosing a good action are different objectives.
Decision-focused (``smart predict-then-optimize'') learning confronts this by training the predictor against the downstream decision loss~\citep{elmachtoub2022spo,mandi2024dfl}, while calibration work aligns predicted scores with the quantities a ranker actually consumes~\citep{guo2017calibration,steck2018calibrated}.
When several decisions must be served at once, recommendation often attaches task-specific heads to a shared representation~\citep{ma2018mmoe,tang2020ple} or shares one predictor across several decision losses~\citep{tang2023mtpo}.
These architectures study parameter sharing across jointly trained objectives.
DSI addresses a complementary question: what should a watch-time estimator expose after it has been trained?
Bayesian decision theory motivates the separation between estimation and task-specific action~\citep{berger1985decision}; DSI operationalizes this principle with lightweight learned readouts.
The contribution is the reusable estimator output; existing multi-task architectures can supply the downstream readout.
This separation also resembles industrial pipelines with distinct scoring, calibration, and reranking stages~\citep{zhao2019recommending,chen2019top}.

\paragraph{Generative recommendation.}
Recent work formulates recommendation as text generation, semantic-ID generation, or autoregressive user-action modeling~\citep{geng2022p5,rajput2023tiger,zhai2024hstu}.
Related systems also generate watch time directly or use a learned watch-time reward to select generated candidates~\citep{deng2025onerec,ma2026gr}.
These methods change the prediction backbone or candidate generator.
DSI instead studies the representation passed from a watch-time estimator to downstream scoring, so the two directions can be combined.


\section{Method}
\label{sec:method}

\begin{figure*}[t]
\centering
\includegraphics[width=\textwidth]{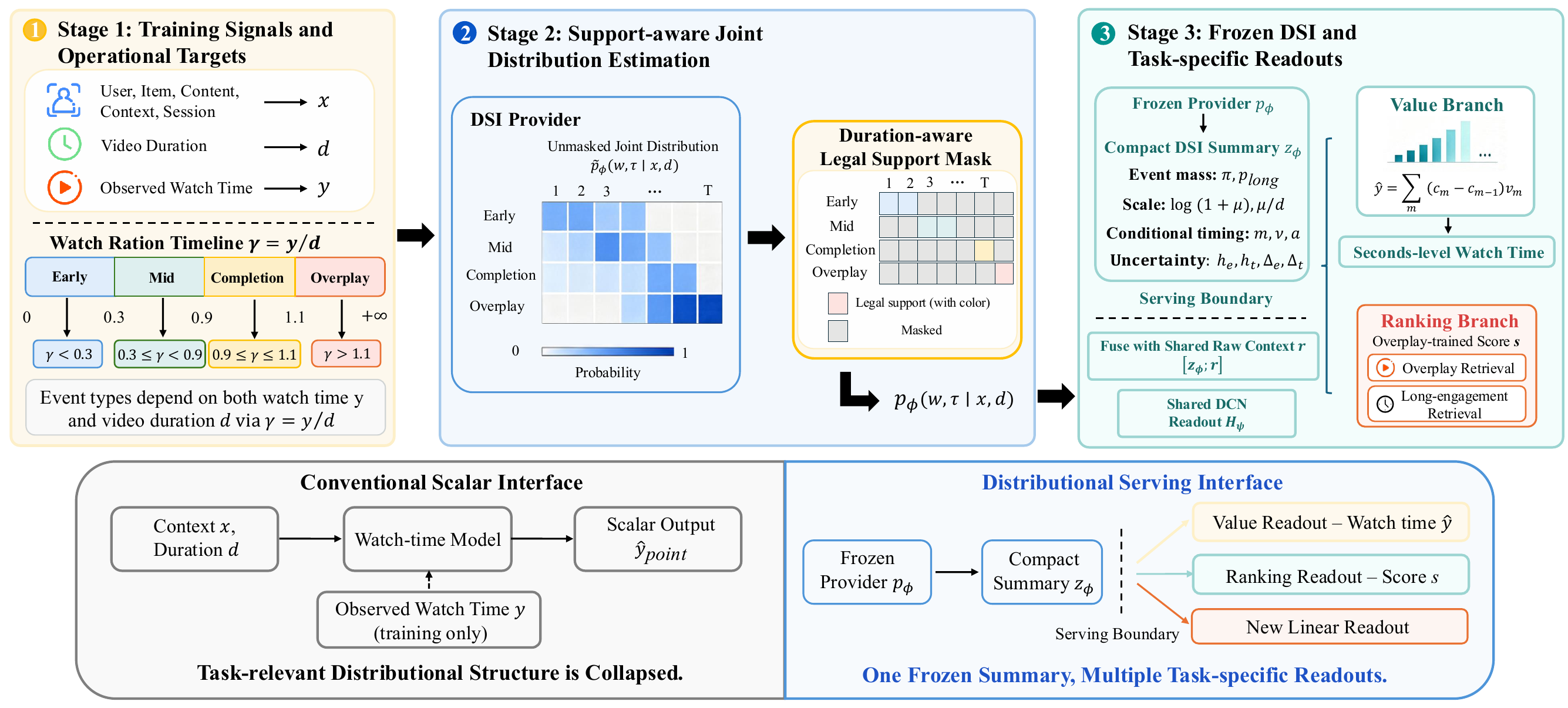}
\caption{The evaluated typed instantiation of DSI comprises three stages. Stage~1 derives four operational event types from watch ratio during training. Stage~2 applies duration-aware support constraints and estimates a joint distribution over event type and event time with watch-time restoration. Stage~3 freezes the provider and exposes its compact summary, the serving-time interface, which a shared DCN readout combines with raw context. The value branch predicts watch time; the ranking branch produces one overplay-trained score for overplay and long-engagement retrieval. The observed watch time supplies training supervision only and is unavailable at inference.}
\Description{Three-stage DSI architecture: operational event-type construction, duration-constrained typed distribution estimation, and a frozen-provider readout with value and ranking branches.}
\label{fig:dsi-architecture}
\end{figure*}

DSI turns one logged watch record into a distributional estimate and a set of task outputs in three stages (Figure~\ref{fig:dsi-architecture}).
Section~\ref{sec:setup} formalizes the data, the limitation of the point interface, and the construction of operational event-type targets from logged watch time and duration.
Section~\ref{sec:provider} trains the \emph{provider}, a joint distribution over event type and event time restricted to duration-dependent support.
Sections~\ref{sec:summary} and~\ref{sec:readout} freeze the provider and pass its compact summary to lightweight task-specific readouts.
Operational event types enter the system in two roles: their targets supervise the provider (Sections~\ref{sec:setup}--\ref{sec:provider}), and their structure organizes the summary exposed at serving time (Section~\ref{sec:summary}). Both roles are instantiation choices of the interface.

\subsection{Problem Setup}
\label{sec:setup}

For each impression $i$, the observed training record is
\begin{equation}
\label{eq:training-record}
\mathcal D=\left\{(\mathbf{x}_i,d_i,y_i)\right\}_{i=1}^{N},
\qquad
\mathbf{x}_i\in\mathcal X,\quad d_i>0,\quad y_i\ge 0,
\end{equation}
where $\mathbf{x}_i$ collects user, item, content, context, and session features, $d_i$ is the video duration, and $y_i$ is the observed watch time in seconds.

At inference time the system does not observe $y_i$, yet it must support decisions that depend on expected seconds watched, on endpoint attainment, and on whether the watch ratio crosses an overplay threshold.
An ordinary watch-time predictor answers all of them with one scalar:
\begin{equation}
\label{eq:point-interface}
\mathbf{q}_i=(\mathbf{x}_i,d_i)
\xrightarrow{\;f_\theta\;}
\hat y_i^{\,\mathrm{point}}\in\mathbb R_{\ge0},
\end{equation}
where $f_\theta$ is the trained predictor and $\hat y_i^{\,\mathrm{point}}$ is its single output.
That output can suffice for a single watch-time objective, and a downstream layer that also reads $d_i$ can turn it into a duration-normalized score.
The pair $(\hat y_i^{\,\mathrm{point}},d_i)$ nonetheless pins down one conditional mean and nothing else, so it leaves open how much probability mass sits near the endpoint and overplay thresholds, and which conditional outcome distribution produced that mean.

Let $\mathcal W=\{\mathrm{early},\mathrm{mid},\mathrm{completion},\mathrm{overplay}\}$ denote the set of operational event types, and let $\tau\in\{1,\ldots,T\}$ index one of $T$ event-time bins.
DSI estimates a joint distribution $p_\phi(w,\tau\mid\mathbf{x},d)$ over event type and event time (Section~\ref{sec:provider}) and exposes it through a summary map $\mathcal G$ and a task-specific readout $\mathcal H_\psi$:
\begin{equation}
\label{eq:decision-interface}
\mathbf{z}_i=\mathcal G\!\left(p_\phi(\cdot,\cdot\mid\mathbf{x}_i,d_i)\right),
\qquad
(\hat y_i,s_i)=\mathcal H_\psi(\mathbf{z}_i,\mathbf{r}_i),
\end{equation}
where $\mathbf{r}_i$ is a fixed subvector of $\mathbf{x}_i$ (the shared raw-context block), $\hat y_i$ is the seconds-level value, and $s_i$ is a ranking score trained for overplay and evaluated for both overplay and long engagement.
The object crossing the serving boundary is thus $\mathbf{z}_i$ rather than $\hat y_i^{\,\mathrm{point}}$: $\mathcal H_\psi$ is the main readout, and a decision added later fits its own readout on the same frozen $\mathbf{z}_i$ (Section~\ref{sec:transfer-probe}).

The operational event-type targets are deterministic functions of the observed record; no new label is collected.
Let $\gamma_i=y_i/d_i$ and let $b(\cdot)\in\{1,\ldots,T\}$ map a nonnegative time to one of $T=69$ upper-edge bins, which use one-second resolution through 60 seconds and progressively wider intervals up to 1800 seconds (exact edges in Appendix~\ref{app:implementation}):
\begin{equation}
\label{eq:operational-target}
(w_i,\tau_i)=
\begin{cases}
(\mathrm{early},b(y_i)), & \gamma_i<\alpha,\\
(\mathrm{mid},b(y_i)), & \alpha\le\gamma_i<\rho,\\
(\mathrm{completion},b(d_i)), & \rho\le\gamma_i\le\beta,\\
(\mathrm{overplay},b(y_i)), & \gamma_i>\beta.
\end{cases}
\end{equation}
The ordered boundaries $0<\alpha<\rho<1<\beta$ are fixed at $\alpha=0.3$, $\rho=0.9$, $\beta=1.1$ before training.
The band $[\rho,\beta]$ gives the endpoint a $10\%$ tolerance, and overplay isolates the upper extreme of the watch-ratio signal. Thresholding watch ratio~\citep{zhao2023d2co,yang2026dadf} and using duration-normalized targets~\citep{zheng2022dvr,zhao2024cwm} are standard practices, motivating the form of our operational event types rather than the exact cutoffs.
Consequently, $w_i$ is an operational event type, not a separately logged action: \emph{overplay} means only $y_i/d_i>\beta$, and these datasets cannot distinguish deliberate replay from automatic looping, cumulative playback, or logging artifacts.
Appendix~\ref{app:external-engagement} examines how these event types associate with separately logged engagement actions.
Appendix~\ref{app:event-boundary} examines the sensitivity of the reported conclusions to the exact boundary values.

\subsection{Typed Event-Time Estimation}
\label{sec:provider}

We call the estimator of $p_\phi$ the \emph{provider} because its frozen output supplies the downstream readouts considered here.
Existing distributional predictors place their distribution over watch time alone~\citep{lin2023tpm,sun2024cread,lin2024cqe,egmn2025}; the object estimated here is joint over event type and event time, and it must respect a structural fact that a time-only distribution never faces: each event type is compatible with only some event times.
For a video of duration $d$, the legal time support of each type is
\begin{equation}
\label{eq:event-support}
\begin{aligned}
\mathcal S_{\mathrm{early}}(d) &= \{\tau:t(\tau)/d<\alpha\},\\
\mathcal S_{\mathrm{mid}}(d) &= \{\tau:\alpha\le t(\tau)/d<\rho\},\\
\mathcal S_{\mathrm{completion}}(d) &= \{b(d)\},\\
\mathcal S_{\mathrm{overplay}}(d) &= \{\tau:t(\tau)/d>\beta\},
\end{aligned}
\end{equation}
where $t(\tau)$ maps a bin to seconds, and the joint legal support is $\mathcal A(d)=\{(w,\tau):\tau\in\mathcal S_w(d)\}$.
The provider outputs one logit per type--time pair, $a_\phi(w,\tau\mid\mathbf{q})$ with $\mathbf{q}=(\mathbf{x},d)$, a grid of $|\mathcal W|\times T$ values, and normalizes them only over this support:
\begin{equation}
\label{eq:support-normalization}
\begin{aligned}
\widetilde p_\phi(w,\tau\mid\mathbf{q})
  &= \frac{\exp a_\phi(w,\tau\mid\mathbf{q})}
  {\sum_{w'\in\mathcal W}\sum_{\tau'=1}^{T}
   \exp a_\phi(w',\tau'\mid\mathbf{q})},\\
p_\phi(w,\tau\mid\mathbf{x},d)
  &= \frac{\widetilde p_\phi(w,\tau\mid\mathbf{q})\,\mathbb 1[(w,\tau)\in\mathcal A(d)]}
  {\sum_{(w',\tau')\in\mathcal A(d)}
   \widetilde p_\phi(w',\tau'\mid\mathbf{q})},
\end{aligned}
\end{equation}
where $\mathbb 1[\cdot]$ is the indicator function, so no probability mass is assigned to incompatible type--time pairs.

Supervision uses the operational event-type target $(w_i,\tau_i)$ of Eq.~\eqref{eq:operational-target}.
Early, mid, and overplay events are supervised at the observed stop bin $b(y_i)$, whereas completion is anchored at the endpoint bin $b(d_i)$.
The endpoint anchor separates completion-band tolerance from time supervision: for a 100-second video, observations of 95, 100, and 108 seconds are all assigned the completion event type and share the coordinate $(\mathrm{completion},b(100))$, because that coordinate represents endpoint attainment rather than the exact logged seconds.
The exact seconds are not discarded; they supervise the restoration term below.
Let
\begin{equation}
\label{eq:time-mean}
\mu(\mathbf{x},d)=\sum_{w\in\mathcal W}\sum_\tau t(\tau)\,p_\phi(w,\tau\mid\mathbf{x},d)
\end{equation}
denote the distributional time mean.
The provider loss combines the support-masked likelihood with log-space restoration:
\begin{equation}
\label{eq:provider-loss}
\begin{aligned}
\ell_{\mathrm{provider}}(i)
&=-\log p_\phi(w_i,\tau_i\mid \mathbf{x}_i,d_i)\\
&\quad+\lambda_{\mathrm{rest}}
\,\mathrm{SmoothL1}\!\left(\log(1+\mu(\mathbf{x}_i,d_i)),\log(1+y_i)\right),
\end{aligned}
\end{equation}
where $\lambda_{\mathrm{rest}}\ge0$ weights the restoration term.
The likelihood term concentrates mass at the observed event-type/time coordinate; the restoration term aligns the log-transformed distributional mean with log-transformed watch time and stabilizes the time statistics consumed by the readout.
On completion records with $y_i\neq d_i$ the two terms compete: the likelihood anchors mass at $b(d_i)$ while restoration pulls $\log(1+\mu)$ toward $\log(1+y_i)$, so part of the off-completion mass compensates watch-time residuals rather than expressing outcome uncertainty.
This competition is bounded by the completion band ($\pm10\%$ of $d$ under $\rho{=}0.9$, $\beta{=}1.1$), and $\lambda_{\mathrm{rest}}$ controls the trade-off.

\subsection{Typed Distribution Summary}
\label{sec:summary}

The full event-time grid holds $|\mathcal W|\times T=276$ numbers per impression, so the readout consumes a compact summary instead.
The summary map $\mathcal G$ of Eq.~\eqref{eq:decision-interface} evaluates every statistic at $(\mathbf{x},d)$ and produces the compact, low-dimensional representation $\mathbf{z}_\phi(\mathbf{x},d)=\mathcal G(p_\phi(\cdot,\cdot\mid\mathbf{x},d))$.
In the typed instantiation evaluated here, this representation has 27 coordinates, and the $\mathbf{z}_i$ in Eq.~\eqref{eq:decision-interface} equals $\mathbf{z}_\phi(\mathbf{x}_i,d_i)$:
\begin{equation}
\label{eq:summary}
\begin{aligned}
\mathbf{z}_\phi(\mathbf{x},d)={}&\big[
\bm\pi,\mathrm{logit}(\bm\pi),p_{\mathrm{long}},
\mathrm{logit}(p_{\mathrm{long}}),\\
&\log(1+\mu),\mu/d,\mathbf{m},\bm\nu,h_e,h_t,
\Delta_e,\Delta_t,\mathbf{a}\big].
\end{aligned}
\end{equation}
Its components derive from the statistics
\begin{equation}
\label{eq:summary-stats}
\begin{gathered}
\begin{aligned}
\pi_w&=\textstyle\sum_{\tau} p_\phi(w,\tau\mid\mathbf{x},d),
&\kappa_\tau&=\textstyle\sum_{w\in\mathcal W} p_\phi(w,\tau\mid\mathbf{x},d),\\
\bar t_w&=\textstyle\pi_w^{-1}\sum_{\tau} t(\tau)\,p_\phi(w,\tau\mid\mathbf{x},d),
&h_e&=-\textstyle\sum_{w\in\mathcal W}\pi_w\log\pi_w,\\
m_w&=\log(1+\bar t_w),
&\nu_w&=\bar t_w/d,
\end{aligned}\\
\mathbf{a}=\big[\nu_{\mathrm{completion}},\nu_{\mathrm{overplay}},\max(\nu_{\mathrm{overplay}}-1,0)\big],
\end{gathered}
\end{equation}
with $\bar t_w=0$ when $\pi_w=0$, and $h_t$ defined on $\bm\kappa=(\kappa_\tau)_{\tau}$ exactly as $h_e$ is on $\bm\pi=(\pi_w)_{w\in\mathcal W}$.
Here $\bm\pi$ and $\bm\kappa$ are the event and time marginals, $p_{\mathrm{long}}=\pi_{\mathrm{completion}}+\pi_{\mathrm{overplay}}$ is the long-watch mass, $\mu$ is the distributional time mean of Eq.~\eqref{eq:time-mean}, $\mathbf{m}=(m_w)_{w\in\mathcal W}$ and $\bm\nu=(\nu_w)_{w\in\mathcal W}$ collect the per-event time scales and ratios, and $\Delta_e$ and $\Delta_t$ are the gaps between the largest and second-largest coordinates of $\bm\pi$ and $\bm\kappa$.
The function $\mathrm{logit}(\cdot)$ acts elementwise on probabilities after numerical clipping.
The event marginals and the ratio vector $\mathbf{a}$ preserve typed structure; the time, entropy, and margin statistics preserve the scale and uncertainty needed for seconds-level error.
The summary is an output of the DSI estimator, not an extra raw feature.

\subsection{Decision Readout}
\label{sec:readout}

Different downstream objectives require different outputs from the same watch log: seconds-level error needs a value estimate, whereas endpoint or overplay retrieval needs a duration-relative ranking score.
DSI therefore freezes the provider and learns lightweight task-specific branches on the fused feature vector $\mathbf{f}_i=[\mathbf{z}_\phi(\mathbf{x}_i,d_i);\mathbf{r}_i]$, with the raw-context block $\mathbf{r}_i$ of Eq.~\eqref{eq:decision-interface}.
Both branches share one parallel DCNv2-style readout parameterized by $\psi$: a value branch for seconds-level prediction and a ranking branch trained for overplay.
The value head predicts bounded ordinal basis scores over data-adaptive thresholds $0=c_0\le c_1\le\cdots\le c_M$:
\begin{equation}
\label{eq:ordinal-head}
v_{im}=\sigma\!\left(g_{\psi,m}^{\mathrm{val}}(\mathbf{f}_i)\right).
\end{equation}
Here $m=1,\ldots,M$, the score $v_{im}\in(0,1)$ weights the interval $(c_{m-1},c_m]$, $g_{\psi,m}^{\mathrm{val}}$ is the $m$-th value output, and $\sigma$ is the logistic sigmoid. The scores are learned independently across thresholds, without a cross-threshold monotonicity constraint.
Seconds-level watch time is recovered by the weighted threshold integral
\begin{equation}
\label{eq:restore}
\hat y_i
=
\sum_{m=1}^{M}
(c_m-c_{m-1}) v_{im},
\qquad c_0=0,
\end{equation}
which bounds $\hat y_i$ to $[0,c_M]$.
Discretize-and-restore value heads of this form are established for watch time~\citep{sun2024cread}, and we keep this one deliberately ordinary: holding the decoder to a known recipe puts the seconds-level comparison on what the head reads rather than on how it decodes.
The ranking branch outputs one event-type score
\begin{equation}
\label{eq:utility-head}
s_i=g_{\psi}^{\mathrm{rank}}(\mathbf{f}_i).
\end{equation}
Here $s_i$ is the ranking logit produced by the shared readout.
Let $u_i\in\{0,1\}$ denote the selected utility label, and let $\ell_{\mathrm{BCE}}(s,u)$ denote binary cross-entropy applied to logit $s$ and label $u$.
The readout objective for impression $i$ is
\begin{align}
\label{eq:readout-loss}
\ell_{\mathrm{readout}}(i)
&=
\lambda_{\mathrm{val}}\,\mathrm{SmoothL1}(\hat y_i,y_i)
+
\lambda_{\mathrm{rank}}\,\ell_{\mathrm{BCE}}(s_i,u_i),
\end{align}
where $\lambda_{\mathrm{val}},\lambda_{\mathrm{rank}}\ge0$ are the value and ranking loss weights.
The first term directly supervises the seconds-level readout, while the second supervises one chosen operational event type.
In the main experiments, $u_i=\mathbb 1[w_i=\mathrm{overplay}]$.
Overplay is the upper extreme of the watch-ratio spectrum, and its support begins only past a duration-dependent boundary, which makes it a demanding duration-relative retrieval probe.
The shared readout checkpoint is selected by validation MAE.
The same overplay-trained score is also evaluated against broader long-engagement relevance, $u_i^{\mathrm{long}}=\mathbb 1[w_i\in\{\mathrm{completion},\mathrm{overplay}\}]$, to test whether the ranking branch transfers beyond the strict overplay event type.
The value branch and ranking branch are identical across all three datasets.
Appendix~\ref{app:implementation} specifies the DCNv2 layers, threshold construction, and optimization settings.
Algorithm~\ref{alg:dsi-training} summarizes the full procedure.

\begin{algorithm}[t]
\caption{DSI Training and Inference}
\label{alg:dsi-training}
\begin{algorithmic}[1]
\Require records $\mathcal D=\{(\mathbf{x}_i,d_i,y_i)\}_{i=1}^{N}$ of Eq.~\eqref{eq:training-record}; boundaries $\alpha,\rho,\beta$; weights $\lambda_{\mathrm{rest}},\lambda_{\mathrm{val}},\lambda_{\mathrm{rank}}$
\Ensure frozen provider $p_\phi$; readout $\mathcal H_\psi$ producing $(\hat y,s)$
\State construct operational event-type targets $(w_i,\tau_i)$ from $(y_i,d_i)$ by Eq.~\eqref{eq:operational-target} \Comment{no new label}
\State train the provider $p_\phi$ of Eq.~\eqref{eq:support-normalization} by minimizing the loss $\ell_{\mathrm{provider}}$ of Eq.~\eqref{eq:provider-loss}
\State freeze $\phi$; compute summaries $\mathbf{z}_i=\mathcal G(p_\phi(\cdot,\cdot\mid\mathbf{x}_i,d_i))$ by Eq.~\eqref{eq:summary}
\State train the readout $\mathcal H_\psi$ on $\mathbf{f}_i=[\mathbf{z}_i;\mathbf{r}_i]$ by minimizing $\ell_{\mathrm{readout}}$ of Eq.~\eqref{eq:readout-loss} \Comment{provider stays fixed}
\State \textbf{inference:} value $\hat y_i$ by Eq.~\eqref{eq:restore}; ranking score $s_i$ by Eq.~\eqref{eq:utility-head}
\end{algorithmic}
\end{algorithm}

\noindent\textbf{Remark.}
Relative-advantage debiasing~\citep{liu2025debiasing} is closest in spirit to DSI: both separate distribution estimation from downstream preference learning.
The two works differ in the problem and in the exposed object.
Relative-advantage debiasing estimates a duration-conditional watch-time distribution for one purpose, converting watch time into a debiased quantile score for ranking; DSI exposes a distributional summary to several readouts with different objectives, including seconds-level value and overplay ranking.


\section{Experiments}
\label{sec:exp}

Our experiments evaluate DSI from predictive performance to interface analysis and reuse.
We first compare the complete system with nine watch time predictors on three public datasets using MAE and XAUC, the two core metrics for watch time prediction, and report retrieval performance on tasks defined relative to video duration.
We then use controls with a shared readout and within one provider to distinguish the contributions of readouts aligned with the target and the distributional information exposed through the serving interface.
Finally, with the provider frozen, we test whether lightweight heads can reuse the same interface for new watch time targets and a separately logged engagement signal.

\subsection{Setup}

\paragraph{Datasets.}
We use three public short-video datasets: KuaiRec and KuaiRand-1K from Kuaishou~\citep{gao2022kuairand,gao2022kuairec}, and WeChat21, a large-scale short-video dataset released by the WeChat Big Data Challenge 2021~\citep{wechat2021challenge}.
We evaluate warm-start prediction with a fixed impression-level random 8:1:1 split.
Past-only features are constructed in user-log order before split assignment, so each impression retains the context available at its original timestamp.
Users and items may appear across splits, and the protocol evaluates within-population prediction without introducing chronological distribution shift.
For WeChat21, we evaluate the duration-below-cap subset ($d<60$ seconds) and remove extreme records with $y/d\ge10$.
All methods use the same processed splits and raw inputs; Appendix~\ref{app:wechat} gives the filtering counts, event-type frequencies, feature construction, and method-specific estimator outputs.
Every reported result uses seeds 2027, 2028, and 2029 for model initialization and training sampling, with the split fixed across seeds.
In the main comparison, DSI's decision readout is trained on the full training split and evaluated on the full test split.
Dataset statistics after preprocessing are given in Table~\ref{tab:datasets} in Appendix~\ref{app:implementation}.

\paragraph{Metrics.}
We report MAE and XAUC for watch-time prediction.
MAE is computed in seconds as $\frac1N\sum_{i=1}^{N}|\hat y_i-y_i|$, where $N$ is the number of evaluated impressions.
XAUC measures pairwise consistency between predicted and observed watch time~\citep{zhan2022d2q,zheng2022dvr}.
For duration-relative retrieval, we report session-level Long-N@5 and OP-N@5.
Long-N@5 uses completion-or-overplay relevance, and OP-N@5 uses overplay relevance; both are derived from the operational event types in Eq.~\eqref{eq:operational-target}, not from logged user actions.
For each session with at least two candidates and positive ideal DCG, we compute NDCG@5 and average the resulting scores equally across eligible sessions; Appendix~\ref{app:metrics} gives the full formula.
Appendix Table~\ref{tab:event-freq} reports the corresponding test-split event-type frequencies.
Lower MAE is better; higher XAUC and retrieval scores are better.
We report the mean and sample standard deviation over three seeds.
Tables boldface the best mean; where a second best is marked, it is underlined.

\paragraph{Baselines.}
We compare against nine watch-time predictors under the same processed splits and evaluation code:
\begin{itemize}
    \item \textbf{VR} regresses raw watch time with mean squared error and serves as the plain point-regression reference used in prior comparisons~\citep{sun2024cread}.
    \item \textbf{WLR}~\citep{covington2016youtube} trains a logistic objective weighted by watch time, following the YouTube ranking recipe.
    \item \textbf{D2Q}~\citep{zhan2022d2q} removes duration bias by regressing duration-group watch-time quantiles.
    \item \textbf{CWM}~\citep{zhao2024cwm} corrects duration bias with a counterfactual watch-time target.
    \item \textbf{DIFL}~\citep{jin2026difl} learns duration-invariant features with a kernel independence penalty and predicts duration-bucket quantiles.
    \item \textbf{TPM}~\citep{lin2023tpm} models watch time with a tree-structured progressive ordinal decomposition.
    \item \textbf{CREAD}~\citep{sun2024cread} discretizes watch time with error-adaptive bins and restores seconds from ordinal classifiers.
    \item \textbf{EGMN}~\citep{egmn2025} fits an exponential-Gaussian mixture over normalized play time.
    \item \textbf{GR}~\citep{ma2026gr} generates watch time autoregressively as a token sequence.
\end{itemize}
Implementation sources, shared adapters, and aligned optimization settings are reported in Appendix~\ref{app:implementation}.
\subsection{Main Results}
\label{sec:main-result}

Table~\ref{tab:main-results} presents the central comparison with nine baselines on the primary watch time prediction metrics, MAE and XAUC.
DSI achieves the lowest MAE on all three datasets, reducing it relative to the strongest baseline by $1.9\%$ to $8.5\%$, and the best XAUC on KuaiRec and KuaiRand-1K; on WeChat21 it ranks second, only $0.0034$ behind CREAD.
The complete DSI system also ranks first on Long-N@5 and OP-N@5 across all three datasets.

\begin{table*}[t]
\centering
\caption{Complete-System Comparison with Nine Baselines on Three Public Datasets}
\label{tab:main-results}
\begin{tabular*}{\textwidth}{@{\extracolsep{\fill}}lrrrr@{}}
\toprule
Method & MAE $\downarrow$ & XAUC $\uparrow$ & Long-N@5 $\uparrow$ & OP-N@5 $\uparrow$ \\
\midrule
\multicolumn{5}{@{}l}{\emph{KuaiRec}} \\
VR    & 4.4518$\pm$0.0236 & 0.5830$\pm$0.0028 & 0.6599$\pm$0.0100 & 0.5799$\pm$0.0075 \\
WLR~\citep{covington2016youtube}   & 4.4652$\pm$0.0663 & 0.6046$\pm$0.0073 & 0.7044$\pm$0.0010 & 0.6572$\pm$0.0009 \\
CWM~\citep{zhao2024cwm}   & 4.3377$\pm$0.0041 & 0.6129$\pm$0.0006 & 0.6863$\pm$0.0005 & 0.6096$\pm$0.0005 \\
EGMN~\citep{egmn2025}  & \underline{3.8327$\pm$0.1617} & 0.6561$\pm$0.0104 & 0.7361$\pm$0.0110 & 0.6863$\pm$0.0122 \\
TPM~\citep{lin2023tpm}   & 3.8757$\pm$0.0059 & 0.6602$\pm$0.0002 & 0.7333$\pm$0.0028 & 0.6829$\pm$0.0035 \\
D2Q~\citep{zhan2022d2q}   & 3.9588$\pm$0.0178 & \underline{0.6605$\pm$0.0010} & \underline{0.7504$\pm$0.0085} & 0.6944$\pm$0.0108 \\
CREAD~\citep{sun2024cread} & 3.8620$\pm$0.2493 & 0.6593$\pm$0.0266 & 0.7484$\pm$0.0119 & \underline{0.6982$\pm$0.0135} \\
GR~\citep{ma2026gr}    & 4.2585$\pm$0.0111 & 0.6023$\pm$0.0011 & 0.6147$\pm$0.0016 & 0.5530$\pm$0.0051 \\
DIFL~\citep{jin2026difl}  & 4.2321$\pm$0.2482 & 0.5887$\pm$0.0728 & 0.7365$\pm$0.0112 & 0.6874$\pm$0.0142 \\
DSI (ours)  & \textbf{3.5059$\pm$0.0044} & \textbf{0.6827$\pm$0.0004} & \textbf{0.8873$\pm$0.0009} & \textbf{0.8808$\pm$0.0001} \\
\midrule
\multicolumn{5}{@{}l}{\emph{KuaiRand-1K}} \\
VR    & 13.8164$\pm$0.1844 & 0.6809$\pm$0.0013 & 0.4992$\pm$0.0067 & 0.3431$\pm$0.0075 \\
WLR~\citep{covington2016youtube}   & 15.0173$\pm$0.2792 & 0.7118$\pm$0.0021 & 0.4540$\pm$0.0005 & 0.3342$\pm$0.0012 \\
CWM~\citep{zhao2024cwm}   & 13.6592$\pm$0.0323 & 0.7052$\pm$0.0005 & 0.5350$\pm$0.0024 & 0.4142$\pm$0.0038 \\
EGMN~\citep{egmn2025}  & 12.8314$\pm$0.1175 & 0.7324$\pm$0.0016 & 0.4608$\pm$0.0044 & 0.3204$\pm$0.0026 \\
TPM~\citep{lin2023tpm}   & 14.7347$\pm$0.1495 & 0.7061$\pm$0.0010 & 0.4796$\pm$0.0053 & 0.3493$\pm$0.0064 \\
D2Q~\citep{zhan2022d2q}   & \underline{12.5668$\pm$0.0251} & \underline{0.7345$\pm$0.0002} & \underline{0.5387$\pm$0.0012} & 0.4035$\pm$0.0019 \\
CREAD~\citep{sun2024cread} & 13.7574$\pm$0.0624 & 0.7192$\pm$0.0021 & 0.4393$\pm$0.0038 & 0.2991$\pm$0.0031 \\
GR~\citep{ma2026gr}    & 15.0652$\pm$0.1533 & 0.6513$\pm$0.0020 & 0.4568$\pm$0.0027 & 0.3320$\pm$0.0038 \\
DIFL~\citep{jin2026difl}  & 13.8023$\pm$0.6272 & 0.6089$\pm$0.0973 & 0.5176$\pm$0.0069 & \underline{0.4311$\pm$0.0608} \\
DSI (ours)  & \textbf{12.3244$\pm$0.0362} & \textbf{0.7367$\pm$0.0010} & \textbf{0.7185$\pm$0.0009} & \textbf{0.7426$\pm$0.0013} \\
\midrule
\multicolumn{5}{@{}l}{\emph{WeChat21}} \\
VR    & 14.1394$\pm$0.0306 & 0.6774$\pm$0.0012 & 0.7357$\pm$0.0009 & 0.6143$\pm$0.0013 \\
WLR~\citep{covington2016youtube}   & 14.0973$\pm$0.1797 & 0.6831$\pm$0.0127 & 0.7397$\pm$0.0031 & 0.6263$\pm$0.0065 \\
CWM~\citep{zhao2024cwm}   & 16.3456$\pm$0.0348 & 0.6625$\pm$0.0003 & \underline{0.7632$\pm$0.0004} & \underline{0.6757$\pm$0.0012} \\
EGMN~\citep{egmn2025}  & 13.4082$\pm$0.3418 & 0.7005$\pm$0.0067 & 0.7472$\pm$0.0055 & 0.6312$\pm$0.0075 \\
TPM~\citep{lin2023tpm}   & 13.6801$\pm$0.0676 & 0.6977$\pm$0.0002 & 0.7503$\pm$0.0019 & 0.6380$\pm$0.0033 \\
D2Q~\citep{zhan2022d2q}   & \underline{13.2964$\pm$0.0301} & 0.6970$\pm$0.0007 & 0.7490$\pm$0.0041 & 0.6380$\pm$0.0069 \\
CREAD~\citep{sun2024cread} & 13.3271$\pm$0.0083 & \textbf{0.7054$\pm$0.0010} & 0.7500$\pm$0.0022 & 0.6338$\pm$0.0030 \\
GR~\citep{ma2026gr}    & 13.9855$\pm$0.0493 & 0.6754$\pm$0.0010 & 0.7322$\pm$0.0007 & 0.6194$\pm$0.0002 \\
DIFL~\citep{jin2026difl}  & 14.8502$\pm$1.8550 & 0.6443$\pm$0.0562 & 0.7300$\pm$0.0219 & 0.6176$\pm$0.0316 \\
DSI (ours)  & \textbf{12.9278$\pm$0.0062} & \underline{0.7020$\pm$0.0004} & \textbf{0.8181$\pm$0.0007} & \textbf{0.8056$\pm$0.0013} \\
\bottomrule
\end{tabular*}
\end{table*}

\begin{figure*}[t]
\centering
\includegraphics[width=\textwidth]{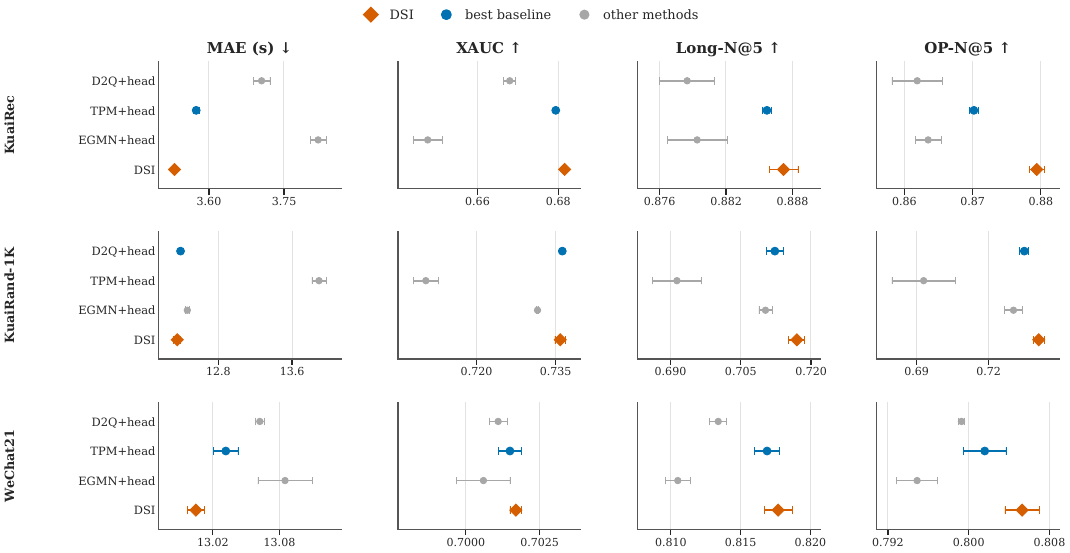}
\caption{Frozen representations are compared under a matched downstream readout. D2Q, TPM, EGMN, and DSI retain their native frozen estimators; each receives the same raw-context bypass, input width, and ordinal-value/overplay-ranking DCN readout trained on the same one million seed-matched impressions, a compute cap applied identically to every estimator. Points show means and bars show sample standard deviations over seeds 2027--2029.}
\Description{Point plots comparing MAE, XAUC, Long-NDCG at 5, and overplay-NDCG at 5 for four frozen representations under an identical readout on three datasets.}
\label{fig:fair-readout}
\end{figure*}

\subsection{Readout and Interface Effects}
\label{sec:matched-readout}

Two complementary controls separate what drives the main-table margins: one varies the exposed representation across providers under a single readout, the other varies only the exposed interface within a single provider.

\paragraph{Across providers, one readout.}
Under a shared readout, task alignment explains most of the absolute retrieval gap, while DSI retains smaller residual gains.
From the nine main-table baselines, we select the three publicly released estimators that expose fixed-dimensional distributional outputs suitable for a frozen representation control: D2Q quantiles, TPM ordinal-tree leaves, and EGMN mixture parameters. These methods span quantile, ordinal, and parametric-mixture representations; the control is not a second ranking of all nine systems.
We freeze D2Q, TPM, EGMN, and DSI and train the same value-plus-overplay readout on each representation (Figure~\ref{fig:fair-readout}).
Each row receives the same raw-context bypass: a baseline method $j$ uses $[\mathbf{r}_i;\mathbf{h}_{ij};\log(1+\widehat y_{ij});\chi(\widehat y_{ij},d_i)]$, where $\mathbf{h}_{ij}$ and $\widehat y_{ij}$ are its frozen native representation and point prediction, and $\chi$ is the shared duration-relative calibration transform.
The baseline input therefore already subsumes the duration-aware point interface $(\widehat y, d, \widehat y/d)$ that a tuple of point prediction and duration would provide.
DSI uses $[\mathbf{r}_i;\mathbf{z}_\phi(\mathbf{x}_i,d_i)]$.
Each provider keeps its native architecture, while the downstream readout, raw context, input width, initialization, and one-million-impression training sample, capped for compute, are fixed across rows.
This control applies the same task-aligned readout to every frozen representation, isolating differences in the exposed representation under matched downstream capacity.

Relative to point-estimate ranking, the matched overplay branch lifts every baseline's retrieval substantially, which places most of the main-table gap in the decision layer rather than in the representation.
DSI still leads in 11 of the 12 dataset--metric cells, with D2Q higher by $0.0004$ XAUC on KuaiRand-1K, but its margins over the strongest baseline representation are far smaller than in Table~\ref{tab:main-results}.
Those residual margins are what separates the representations once every estimator has the same decision layer.

\paragraph{Within-provider interface control.}
This control tests whether a distributional interface adds useful information beyond a duration-aware point interface that already includes $\mu/d$ (Table~\ref{tab:interface-width}).
It fixes the provider, readout, raw-context bypass, and the one-million-impression sample used to train the readout, and varies only the representation served across the boundary: the duration-aware mean summary $[\log(1+\mu),\mu/d]$ computed from Eq.~\eqref{eq:time-mean}, the compact low-dimensional summary of Eq.~\eqref{eq:summary} (27 dimensions), or the full 69-bin time marginal.
The compact summary beats the mean summary on MAE, XAUC, and OP-N@5 on all three datasets, though the margins clear seed variation only on KuaiRec and WeChat21, and Long-N@5 reverses on two of the three.
That the one uniformly signed retrieval gain is OP-N@5 follows the design: the summary encodes the overplay region explicitly through $\mathbf{a}$, whereas a duration-aware mean can only locate that region through $\mu/d$.
The full 69-bin marginal is the informative case, matching the compact summary on MAE while giving the weakest retrieval of the three interfaces, since integrating out the event-type axis leaves time resolution without outcome structure.
These results show that the compact distributional interface preserves predictive information beyond a duration-aware mean and makes it available to downstream readouts.

\begin{table}[b!]
\centering
\caption{Interface Comparison Within One Frozen Provider}
\label{tab:interface-width}
\footnotesize
\setlength{\tabcolsep}{1.5pt}
\renewcommand{\arraystretch}{1.06}
\begin{tabular*}{\columnwidth}{@{\extracolsep{\fill}}lrrrr@{}}
\toprule
Interface & MAE $\downarrow$ & XAUC $\uparrow$ & Long-N@5 $\uparrow$ & OP-N@5 $\uparrow$ \\
\midrule
\multicolumn{5}{@{}l}{\emph{KuaiRec}} \\
Mean+ratio & $3.6063{\pm}0.0039$ & $0.6763{\pm}0.0015$ & $\mathbf{0.8899{\pm}0.0004}$ & $0.8752{\pm}0.0004$ \\
Compact-27 & $3.5309{\pm}0.0003$ & $0.6820{\pm}0.0003$ & $0.8880{\pm}0.0011$ & $\mathbf{0.8792{\pm}0.0003}$ \\
Full-69 & $\mathbf{3.5161{\pm}0.0014}$ & $\mathbf{0.6822{\pm}0.0004}$ & $0.8867{\pm}0.0006$ & $0.8725{\pm}0.0003$ \\
\midrule
\multicolumn{5}{@{}l}{\emph{KuaiRand-1K}} \\
Mean+ratio & $12.4029{\pm}0.0696$ & $0.7357{\pm}0.0009$ & $\mathbf{0.7181{\pm}0.0002}$ & $0.7409{\pm}0.0013$ \\
Compact-27 & $12.3820{\pm}0.0823$ & $\mathbf{0.7362{\pm}0.0005}$ & $0.7178{\pm}0.0006$ & $\mathbf{0.7417{\pm}0.0004}$ \\
Full-69 & $\mathbf{12.3646{\pm}0.0461}$ & $0.7360{\pm}0.0008$ & $0.7078{\pm}0.0020$ & $0.7280{\pm}0.0020$ \\
\midrule
\multicolumn{5}{@{}l}{\emph{WeChat21}} \\
Mean+ratio & $13.1278{\pm}0.0070$ & $0.7012{\pm}0.0002$ & $0.8149{\pm}0.0005$ & $0.8019{\pm}0.0005$ \\
Compact-27 & $\mathbf{13.0196{\pm}0.0035}$ & $\mathbf{0.7014{\pm}0.0003}$ & $\mathbf{0.8165{\pm}0.0004}$ & $\mathbf{0.8035{\pm}0.0005}$ \\
Full-69 & $13.0379{\pm}0.0082$ & $0.7007{\pm}0.0006$ & $0.8129{\pm}0.0005$ & $0.7981{\pm}0.0006$ \\
\bottomrule
\end{tabular*}
\par\smallskip
\begin{minipage}{\columnwidth}
\scriptsize
Only the exposed interface changes; provider, readout, raw-context bypass, and training sample are fixed. Bold marks the best mean within each dataset over seeds 2027--2029; with three interfaces we mark no second-best. All readouts in this control use the same training sample of one million impressions, whereas Table~\ref{tab:main-results} uses the full training split.
\end{minipage}
\end{table}

\paragraph{The symmetric control.}
The complementary direction holds the interface fixed and varies the parameterization: a time-only provider and a plain 27-dimensional summary of matched dimension each perform on par with the typed instantiation on the main and transfer targets (Appendix~\ref{app:time-only}).
Together, these controls associate the remaining differences with what the readout receives, while showing that the tested provider and summary parameterizations perform similarly under a fixed interface.
We use the typed instantiation as our default for its event-level interpretation and duration-consistent support.
Comparable results from the time-only provider and plain summary show that the same serving interface works with less structured parameterizations as well.
Component-level mechanism ablations, covering the input representation, the readout branches, restoration, and the event-boundary values, appear in Appendix~\ref{app:mechanism}.
\subsection{Reusability of the Frozen Interface}
\label{sec:transfer-probe}

The preceding controls evaluate the original DSI targets.
We next freeze the providers and fit linear heads for two kinds of new targets: targets defined from the same watch-time observations but absent from the original objectives, and an engagement label recorded separately from watch time and unused during provider training.
No provider is retrained.

\paragraph{New watch-time targets.}
On KuaiRec, we define two targets from the watch-time observations used to train the providers but absent from the original decision readouts: a binary watch-ratio target $\mathbb 1[y_i/d_i>0.5]$, whose threshold falls inside the mid-watch interval rather than on an event boundary, and the conditional $0.75$ quantile of watch time, trained with pinball loss in $\log(1+y_i)$ space.
Each provider exposes its native distributional output, DSI its compact summary, TPM its 32-leaf distribution, CREAD its 50-dimensional ordinal survival scores, EGMN its mixture parameters, and D2Q its quantile and duration-relative outputs; all are right-padded to 82 dimensions, receive the same clean raw-context bypass, and feed the same linear head.
Every head trains for 6,000 updates on seed-matched samples of $1\%$ or $100\%$ of the training labels, with checkpoints selected on the full validation split.
Because the $1\%$ condition still holds 97,779 impressions, it tests a relative label budget rather than a few-shot regime.

\begin{table}[tb!]
\centering
\caption{Frozen-Interface Transfer to New Watch-Time Targets on KuaiRec}
\label{tab:transfer-probe}
\footnotesize
\setlength{\tabcolsep}{1.5pt}
\renewcommand{\arraystretch}{1.06}
\begin{tabular*}{\columnwidth}{@{\extracolsep{\fill}}lrrrr@{}}
\toprule
Representation & Ratio-AUC & Ratio-N@5 & Q75 Pinball & Q75 MAE \\
\midrule
\multicolumn{5}{@{}l}{\emph{1\% labels}} \\
D2Q   & $0.8354{\pm}0.0006$ & $0.9252{\pm}0.0002$ & $0.1682{\pm}0.0007$ & $4.507{\pm}0.028$ \\
TPM   & $0.8406{\pm}0.0008$ & $0.9294{\pm}0.0003$ & $\underline{0.1524{\pm}0.0003}$ & $\underline{4.471{\pm}0.015}$ \\
CREAD & $\underline{0.8746{\pm}0.0152}$ & $\underline{0.9307{\pm}0.0058}$ & $0.1525{\pm}0.0096$ & $4.481{\pm}0.233$ \\
EGMN  & $0.8244{\pm}0.0078$ & $0.9227{\pm}0.0016$ & $0.1635{\pm}0.0016$ & $4.882{\pm}0.083$ \\
DSI (ours)  & $\mathbf{0.8852{\pm}0.0005}$ & $\mathbf{0.9324{\pm}0.0001}$ & $\mathbf{0.1488{\pm}0.0004}$ & $\mathbf{4.310{\pm}0.015}$ \\
\midrule
\multicolumn{5}{@{}l}{\emph{100\% labels}} \\
D2Q   & $0.8355{\pm}0.0005$ & $0.9251{\pm}0.0003$ & $0.1684{\pm}0.0007$ & $4.548{\pm}0.077$ \\
TPM   & $0.8401{\pm}0.0009$ & $0.9293{\pm}0.0009$ & $\underline{0.1536{\pm}0.0005}$ & $\underline{4.527{\pm}0.065}$ \\
CREAD & $\underline{0.8743{\pm}0.0152}$ & $\underline{0.9302{\pm}0.0058}$ & $0.1544{\pm}0.0101$ & $4.620{\pm}0.313$ \\
EGMN  & $0.8243{\pm}0.0068$ & $0.9227{\pm}0.0009$ & $0.1641{\pm}0.0016$ & $5.120{\pm}0.218$ \\
DSI (ours)  & $\mathbf{0.8849{\pm}0.0006}$ & $\mathbf{0.9327{\pm}0.0001}$ & $\mathbf{0.1494{\pm}0.0003}$ & $\mathbf{4.331{\pm}0.077}$ \\
\bottomrule
\end{tabular*}
\end{table}

DSI has the best mean on all four metrics at both label fractions, and with $1\%$ of the labels it already beats the strongest full-label baseline on each of them (Table~\ref{tab:transfer-probe}).
Its Q75 margins exceed seed variation, while its ratio gains are smaller.
Performance changes little between $1\%$ and $100\%$ labels, indicating that the linear heads saturate early under the fixed-update protocol.

\paragraph{A separately logged engagement target.}
The first transfer test remains within watch-time supervision.
We therefore test whether the frozen DSI interface transfers to a feedback signal recorded separately from watch time.
On KuaiRand-1K, an impression is positive when the user likes, follows, comments, or forwards the video.
This label is not used to train the provider and is not included in the raw-context bypass.
This test reuses the frozen DSI provider of Table~\ref{tab:main-results} and fits a linear head with binary cross-entropy on $1\%$, $10\%$, or $100\%$ of the training labels, under the first step's update budget and selection rule.
We compare the learned DSI summary with the bypass alone and with the same summary from a randomly initialized provider.
The random-provider control preserves the additional feature path while removing learned watch-time structure.

\begin{table}[tb!]
\centering
\caption{Transfer to a Separately Logged Engagement Target on KuaiRand-1K}
\label{tab:engagement-probe}
\footnotesize
\setlength{\tabcolsep}{1.5pt}
\renewcommand{\arraystretch}{1.06}
\begin{tabular*}{\columnwidth}{@{\extracolsep{\fill}}lrrr@{}}
\toprule
Labels & Bypass & +Random summary & +DSI summary \\
\midrule
\multicolumn{4}{@{}l}{\emph{AUC}} \\
$1\%$   & $0.6678{\pm}0.0004$ & $0.6658{\pm}0.0031$ & $\mathbf{0.6920{\pm}0.0030}$ \\
$10\%$  & $0.6678{\pm}0.0005$ & $0.6650{\pm}0.0029$ & $\mathbf{0.6903{\pm}0.0015}$ \\
$100\%$ & $0.6669{\pm}0.0011$ & $0.6630{\pm}0.0008$ & $\mathbf{0.6882{\pm}0.0013}$ \\
\midrule
\multicolumn{4}{@{}l}{\emph{N@5}} \\
$1\%$   & $0.4266{\pm}0.0016$ & $0.4292{\pm}0.0033$ & $\mathbf{0.4362{\pm}0.0032}$ \\
$10\%$  & $0.4270{\pm}0.0018$ & $0.4264{\pm}0.0007$ & $\mathbf{0.4336{\pm}0.0059}$ \\
$100\%$ & $0.4257{\pm}0.0007$ & $0.4279{\pm}0.0045$ & $\mathbf{0.4328{\pm}0.0083}$ \\
\bottomrule
\end{tabular*}
\end{table}

Adding the frozen summary to the bypass improves AUC by $0.021$--$0.024$ at every label budget, with the $1\%$ result comparable to the full-label one, and NDCG@5 moves the same way though only its $1\%$ gain clears the seed spread (Table~\ref{tab:engagement-probe}).
The random-provider summary performs on par with or below the bypass alone, showing that the gain comes from learned watch-time structure rather than an extra input channel.
Together, the two transfer tests show that the frozen interface can be reused for new watch-time targets and a separately logged engagement target by training only one linear head per target.

\section{Conclusion}
\label{sec:conclusion}

We presented DSI, a framework with three stages that turns a structured distribution of watch outcomes into a compact serving interface for lightweight downstream readouts.
Across three public datasets collected from short video platforms, DSI achieves the lowest MAE among nine baselines on all three and the best XAUC on two.
With the provider and readout fixed, its compact summary further improves MAE and XAUC over a mean paired with video duration, confirming that the serving interface preserves predictive information beyond a scalar estimate.
Readouts trained for each task also convert the same estimate into strong rankings for long engagement and overplay.
Once frozen, DSI supports a new threshold on watch ratio, a conditional upper quantile, and a separately logged engagement signal through linear heads alone.
DSI therefore turns accurate watch time estimation into a reusable decision service that supports multiple recommendation objectives with one frozen estimator and a common distributional interface.

\bibliographystyle{ACM-Reference-Format}
\bibliography{references}


\clearpage
\appendix

\section{Implementation Details}
\label{app:implementation}

\begin{table}[H]
\centering
\caption{Dataset Statistics After Impression-Level Preprocessing}
\label{tab:datasets}
\footnotesize
\renewcommand{\arraystretch}{1.06}
\begin{tabular*}{\columnwidth}{@{\extracolsep{\fill}}lrrrrr@{}}
\toprule
Dataset & Users & Items & Train & Val & Test \\
\midrule
KuaiRec & 7,176 & 10.7K & 9.8M & 1.2M & 1.2M \\
KuaiRand-1K & 1,000 & 3.8M & 8.6M & 1.1M & 1.1M \\
WeChat21 & 20,000 & 78.9K & 4.8M & 0.6M & 0.6M \\
\bottomrule
\end{tabular*}
\end{table}

\paragraph{Provider.}
The typed event-time provider uses 69 upper-edge time bins:
\begin{equation}
\begin{aligned}
\mathbf{u}&=(1,\ldots,60,90,135,203,305,458,\\
&\hspace{2.7em}687,1031,1547,1800),\\
u_{k+1}&=\min(1800,\lceil1.5u_k\rceil)\quad (u_k\ge60).
\end{aligned}
\end{equation}
For any nonnegative time $t$,
\begin{equation}
\begin{aligned}
b(t)&=\min\{k:u_k\ge\operatorname{clip}(\lceil t\rceil,1,1800)\},\\
t(k)&=u_k.
\end{aligned}
\end{equation}
Values above 1800 seconds map to the final bin rather than being discarded. The same upper-edge vector is used by the support mask, completion anchor, and restoration.
The provider backbone uses embedding dimension 32, hidden dimension 128, and dropout 0.1. Dense parameters use AdamW with weight decay $10^{-6}$; WeChat21's sparse embeddings use SparseAdam. The learning rate is $10^{-3}$ for KuaiRec and KuaiRand-1K and $10^{-2}$ for WeChat21, including its sparse optimizer. Training has no learning-rate scheduler or warmup, clips the global gradient norm at 10, runs for at most five epochs, and stops after two validation epochs without improvement. The provider checkpoint minimizes validation exact joint event--time NLL; event-marginal calibration is fitted on the validation split only after that checkpoint is loaded. We use $\lambda_{\mathrm{rest}}=1$, with PyTorch's default SmoothL1 transition parameter of $1$ in $\log(1+t)$ units.

\paragraph{Decision readout.}
The decision readout is a parallel DCNv2-style backbone~\citep{wang2021dcnv2} with two full-rank cross layers and a two-layer deep branch of widths $[128,128]$; each deep layer uses LayerNorm, GELU, and dropout 0.1, and the cross and deep outputs are concatenated before the task branches.
This is the matrix-parameterized DCNv2 form rather than the vector-parameterized original DCN layer.
Writing $\mathbf{o}_0=\mathbf{f}_i$ for the fused input of Section~\ref{sec:readout}, the cross branch applies
\begin{equation}
\mathbf{o}_{l+1}=\mathbf{o}_l+\mathbf{o}_0\odot\operatorname{Dropout}(\mathbf{W}_l\mathbf{o}_l+\mathbf{b}_l),
\qquad \mathbf{o}_{\mathrm{cross}}=\operatorname{LayerNorm}(\mathbf{o}_L),
\end{equation}
where each $\mathbf{W}_l\in\mathbb{R}^{D\times D}$ is full rank; the resulting cross vector is concatenated with the deep-branch output.
The readout uses batch size 8192, AdamW, learning rate $10^{-3}$, weight decay $10^{-4}$, and gradient clipping at 10. It has no scheduler, warmup, or early stopping; all five epochs are trained and the checkpoint with the lowest validation MAE is retained.
The ordinal branch uses $M=30$ thresholds and $\lambda_{\mathrm{val}}=30$; the ranking branch uses $\lambda_{\mathrm{rank}}=0.5$.
The 30 thresholds are obtained from training-set watch-time quantiles. Candidate quantile levels use the error-adaptive warp of CREAD~\citep{sun2024cread}, $q_m(a)=(1-e^{-a m/M})/(1-e^{-a})$, with $q_m(0)=m/M$; a one-dimensional training-set search selects $a$ with threshold-search weight $3$, and $c_M$ is the maximum training watch time. The value branch combines the resulting ordinal basis scores through Eq.~\eqref{eq:restore}, which bounds $\hat y_i$ to $[0,c_M]$, and applies SmoothL1 supervision with PyTorch's default transition parameter of $1$ second. The ranking branch uses binary cross-entropy on the overplay event-type target.
The main protocol uses this DCNv2-style readout on fused raw-context and typed-summary features.

\paragraph{Baselines.}
All baselines use the same processed loader and full training split. They use batch size 4096, weight decay $10^{-6}$, gradient clipping at 10, no scheduler or warmup, and five fixed epochs. The learning rate is $10^{-2}$ except for GR, for which a small learning-rate sweep selected $5\times10^{-3}$ after $10^{-2}$ produced unstable utility metrics. VR, WLR, D2Q, TPM, CREAD, and EGMN share the DADF-style field adapter~\citep{yang2026dadf}: categorical fields use 16-dimensional embeddings, continuous fields use learned scalar projections, sequence fields use masked mean pooling, and the common MLP width is $[256,128,64]$. CWM, DIFL, and GR retain their method-specific FM, invariant-feature, and encoder--decoder backbones. Table~\ref{tab:baseline-implementation} records the implementation source and main method-specific settings.

\begin{table}[H]
\centering
\caption{Baseline Implementation Sources and Key Settings}
\label{tab:baseline-implementation}
\scriptsize
\renewcommand{\arraystretch}{1.04}
\begin{tabular}{@{}p{0.12\columnwidth}p{0.25\columnwidth}p{0.55\columnwidth}@{}}
\toprule
Method & Source & Main method-specific setting \\
\midrule
VR & Paper reproduction & Wide\&Deep; embedding 16; MLP $[256,128,64]$ \\
WLR & Paper reproduction & Same backbone as VR; watch-time-weighted logistic loss \\
CWM & Authors' code (\texttt{c36da4b}) & FM backbone; embedding 10; $C=40$, $\sigma=2$ \\
D2Q & Authors' code (\texttt{4b6e68c}) & 30 duration buckets; 101 empirical quantile points \\
DIFL & Authors' code (\texttt{4974910}) & Feature width 32; 30 duration buckets; 101 quantile points \\
TPM & Authors' code (\texttt{eaa2f5d}) & 32-leaf progressive ordinal tree \\
CREAD & Paper reproduction & 50 ordinal heads; search weight 50; head MLP $[32]$ \\
EGMN & Authors' code (\texttt{162b09b}) & One exponential and 10 Gaussian components; dropout 0.2 \\
GR & Authors' code (\texttt{0cd3593}) & Hidden width 128; 8 heads; 3 decoder layers; FFN 256 \\
\bottomrule
\end{tabular}
\par\smallskip
\begin{minipage}{\columnwidth}
\scriptsize
Public implementations were adapted only at the data interface. The shared field adapter follows the public DADF code at commit \texttt{033291d}; VR, WLR, and CREAD have no dedicated source repository in our benchmark package and are reproduced from their papers.
\end{minipage}
\end{table}

\section{Metric Definitions}
\label{app:metrics}

The top-$k$ retrieval metrics adapt NDCG-style ranking evaluation to relevance derived from the operational event types. We use them to study task-specific ranking; they are not standardized benchmark metrics.
XAUC is a global pairwise continuous AUC over watch time.
Long-N@5 uses completion-or-overplay relevance.
OP-N@5 uses overplay relevance $\mathbb 1[w_i=\mathrm{overplay}]=\mathbb 1[y_i/d_i>\beta]$.
For a session $s$ with candidate set $\mathcal I_s$, the method score ranks its impressions and gives the $j$-th ranked impression relevance $r_{s,j}$.
We compute
\begin{equation}
\begin{aligned}
\operatorname{DCG@5}(s)
&=\sum_{j=1}^{\min(5,|\mathcal I_s|)}\frac{r_{s,j}}{\log_2(j+1)},\\
\operatorname{N@5}(s)
&=\frac{\operatorname{DCG@5}(s)}{\operatorname{IDCG@5}(s)},
\end{aligned}
\end{equation}
where $\operatorname{IDCG@5}(s)$ uses the ideal relevance ordering.
Long-N@5 and OP-N@5 are the equal-weight means of $\operatorname{N@5}(s)$ over sessions with at least two candidates and positive ideal DCG.
DSI uses its overplay ranking score, while the standalone baselines use predicted watch time; matched-readout controls use the common readout score.
Table~\ref{tab:event-freq} reports the test-split frequency of the four operational event types; the overplay column is the prevalence of OP-N@5 relevance under Eq.~\eqref{eq:operational-target}.

\begin{table}[H]
\centering
\caption{Operational Event-Type Frequencies in the Test Split (\%)}
\label{tab:event-freq}
\footnotesize
\renewcommand{\arraystretch}{1.06}
\begin{tabular*}{\columnwidth}{@{\extracolsep{\fill}}lrrrr@{}}
\toprule
Dataset & Early & Mid & Compl. & Overplay \\
\midrule
KuaiRec & 22.5 & 37.3 & 11.2 & 29.0 \\
KuaiRand-1K & 65.1 & 15.5 & 9.1 & 10.3 \\
WeChat21 & 36.9 & 14.1 & 12.5 & 36.5 \\
\bottomrule
\end{tabular*}
\end{table}
The ablations also report Long-AUC and OP-AUC.
WT-N@5 uses $\log(1+y_i)$ relevance and does not depend on the event-type boundaries.

\section{Mechanism Ablations}
\label{app:mechanism}

\begin{figure*}[tb]
\centering
\includegraphics[width=\textwidth]{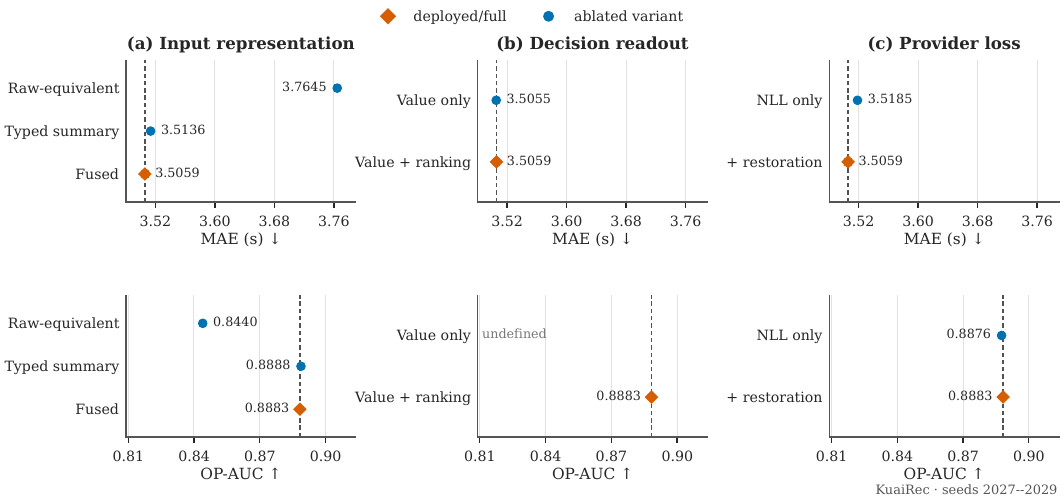}
\caption{Mechanism ablations isolate the roles of the typed summary, ranking branch, and restoration on KuaiRec. Points show three-seed means and bars show sample standard deviations. (a) The typed summary carries most of the fused readout signal. (b) The ranking branch supplies OP-AUC without changing MAE; OP-AUC is undefined for the value-only branch. (c) Watch-time restoration improves MAE while leaving OP-AUC nearly unchanged.}
\Description{Six point plots showing MAE and overplay AUC for input representation, decision-readout, and provider-loss ablations.}
\label{fig:mechanism-ablation}
\end{figure*}

The ablations assign the gains to three distinct components: the exposed summary carries the input signal, the ranking branch produces retrieval, and restoration improves seconds-level prediction.
Figure~\ref{fig:mechanism-ablation} reports three KuaiRec ablations under the corresponding controlled settings.
The provider and decision readout use the full training split in all three controls.

\paragraph{Input representation.}
The typed summary retains most of the readout signal.
Holding the readout fixed, we compare raw, typed, and fused inputs with the same 43-dimensional layout and initialization.
Their inputs are $[\mathbf{r}_i;\mathbf{0}]$, $[\mathbf{0};\mathbf{z}_\phi(\mathbf{x}_i,d_i)]$, and $[\mathbf{r}_i;\mathbf{z}_\phi(\mathbf{x}_i,d_i)]$, respectively.
The typed input is stronger than the raw input (MAE $3.5136$ vs.\ $3.7645$, OP-AUC $0.8888$ vs.\ $0.8440$) and remains close to the fused input.
The raw-context bypass lowers MAE from $3.5136$ to $3.5059$ but gives no consistent retrieval gain.
The raw input already reaches OP-AUC $0.844$ with the task-matched head, showing that the head supplies most of the absolute retrieval lift; the typed summary supplies the residual gain to $0.889$.

\paragraph{Readout branches.}
The ranking branch changes retrieval but not point estimation.
Adding the overplay-ranking branch changes MAE by only $+0.0005\pm0.0031$ seconds and XAUC by less than $10^{-4}$ on average.
The value-only model has no typed ranking output; the trained ranking branch reaches OP-AUC $0.8883$ and OP-N@5 $0.8808$.

\paragraph{Restoration.}
Restoration improves point estimation while leaving retrieval almost unchanged, which also quantifies the likelihood--restoration competition noted in Section~\ref{sec:provider}.
Adding restoration to the support-masked likelihood lowers MAE from $3.5185$ to $3.5059$ and raises XAUC from $0.6803$ to $0.6827$.
The paired MAE reduction is $0.0125\pm0.0029$ seconds, and all three seed-wise differences have the same sign.
OP-AUC changes by less than $0.001$, so restoration affects point estimation rather than retrieval.

\section{Event-Boundary Sensitivity}
\label{app:event-boundary}

The event boundaries assign records to the operational event types.
They are not tuned on the reported test metrics.
Every row retrains the provider and readout on the full training split with a single seed (2027). The table is a within-sweep sensitivity analysis; the main table reports the separate three-seed protocol.
Table~\ref{tab:event-boundary-sensitivity} sweeps one boundary at a time on all three datasets while keeping the others at the default $\alpha=0.3,\rho=0.9,\beta=1.1$.
The label-independent metrics, seconds-level MAE and WT-N@5, remain stable across these perturbations.
The $\alpha$ sweep leaves both retrieval labels fixed; the $\rho$ sweep redefines the long-engagement label, and the $\beta$ sweep redefines the overplay label.
Long-AUC under $\rho$ and OP-AUC under $\beta$ therefore measure target-definition sensitivity rather than a change in model capacity.

\begin{table}[H]
\centering
\caption{Event-Boundary Sensitivity Across Three Datasets Under the Full-Training Protocol}
\label{tab:event-boundary-sensitivity}
\scriptsize
\setlength{\tabcolsep}{1pt}
\renewcommand{\arraystretch}{1.02}
\begin{tabular*}{\columnwidth}{@{\extracolsep{\fill}}llrrrr@{}}
\toprule
Boundary & Value & MAE & Long-AUC & OP-AUC & WT-N@5 \\
\midrule
\multicolumn{6}{@{}l}{\emph{KuaiRec}} \\
$\alpha$ & 0.20 & 3.5428 & 0.8361 & 0.8843 & 0.9383 \\
$\alpha$ & 0.30 & 3.5362 & 0.8361 & 0.8851 & 0.9382 \\
$\alpha$ & 0.40 & 3.5348 & 0.8354 & 0.8846 & 0.9382 \\
$\rho$ & 0.80 & 3.5297 & 0.8074 & 0.8849 & 0.9385 \\
$\rho$ & 0.90 & 3.5362 & 0.8361 & 0.8851 & 0.9382 \\
$\rho$ & 0.95 & 3.5302 & 0.8518 & 0.8846 & 0.9388 \\
$\beta$ & 1.05 & 3.5309 & 0.8494 & 0.8830 & 0.9385 \\
$\beta$ & 1.10 & 3.5362 & 0.8361 & 0.8851 & 0.9382 \\
$\beta$ & 1.20 & 3.5350 & 0.8063 & 0.8902 & 0.9385 \\
\midrule
\multicolumn{6}{@{}l}{\emph{KuaiRand-1K}} \\
$\alpha$ & 0.20 & 12.3720 & 0.8133 & 0.8889 & 0.8016 \\
$\alpha$ & 0.30 & 12.3819 & 0.8115 & 0.8889 & 0.8010 \\
$\alpha$ & 0.40 & 12.3841 & 0.8141 & 0.8891 & 0.8012 \\
$\rho$ & 0.80 & 12.3833 & 0.8105 & 0.8891 & 0.8015 \\
$\rho$ & 0.90 & 12.3819 & 0.8115 & 0.8889 & 0.8010 \\
$\rho$ & 0.95 & 12.3769 & 0.8147 & 0.8889 & 0.8018 \\
$\beta$ & 1.05 & 12.3840 & 0.8184 & 0.8808 & 0.8019 \\
$\beta$ & 1.10 & 12.3819 & 0.8115 & 0.8889 & 0.8010 \\
$\beta$ & 1.20 & 12.3766 & 0.8049 & 0.8919 & 0.8013 \\
\midrule
\multicolumn{6}{@{}l}{\emph{WeChat21}} \\
$\alpha$ & 0.20 & 13.1313 & 0.7523 & 0.7880 & 0.8619 \\
$\alpha$ & 0.30 & 13.1414 & 0.7517 & 0.7875 & 0.8624 \\
$\alpha$ & 0.40 & 13.1308 & 0.7533 & 0.7877 & 0.8626 \\
$\rho$ & 0.80 & 13.1271 & 0.7531 & 0.7880 & 0.8620 \\
$\rho$ & 0.90 & 13.1414 & 0.7517 & 0.7875 & 0.8624 \\
$\rho$ & 0.95 & 13.1302 & 0.7551 & 0.7881 & 0.8629 \\
$\beta$ & 1.05 & 13.1514 & 0.7638 & 0.7765 & 0.8613 \\
$\beta$ & 1.10 & 13.1414 & 0.7517 & 0.7875 & 0.8624 \\
$\beta$ & 1.20 & 13.1193 & 0.7435 & 0.7911 & 0.8629 \\
\bottomrule
\end{tabular*}
\vspace{2pt}
\scriptsize MAE and WT-N@5 do not depend on the boundaries between operational event types. The $\rho$ sweep changes long-engagement relevance, and the $\beta$ sweep changes overplay relevance; their corresponding AUC movements are target-definition sensitivity, not fixed-label capacity.
\end{table}

\section{Interface Robustness Controls}
\label{app:time-only}

This appendix is the fixed-interface half of the symmetric attribution of Section~\ref{sec:matched-readout}: each control replaces one part of the typed parameterization while the exposed interface and the rest of the pipeline stay fixed. If the gains are primarily interface-level, performance should remain comparable.
Three controls replace, respectively, the provider supervision on the main tasks, the provider supervision on the transfer targets, and the form of the exposed summary.
\FloatBarrier

\paragraph{Provider supervision.}
This control removes typed supervision, support masking, and completion anchoring from provider training while keeping everything else fixed.
Both providers share the input encoder, hidden width, time bins, $4\times T$ output budget, restoration term, training data, seeds, and optimizer.
The time-only provider averages the four logit groups along the event axis into one ordinary $T$-bin distribution $p_\phi(\tau\mid\mathbf{x},d)$ and is trained with categorical NLL at the observed bin $b(y_i)$, with no operational event-type targets, support mask, or completion anchor.
Its distribution enters the same downstream pipeline through the same summary map: the four event masses are obtained by region integration, $\eta_w(\tau)=p_\phi(\tau\mid\mathbf{x},d)\,\mathbb 1[t(\tau)/d\in I_w]$, where $I_w$ are the watch-ratio intervals of Eq.~\eqref{eq:operational-target}.
Both frozen providers feed the identical DCN readout trained on one million seed-matched impressions.

The two providers are close on all datasets.
The small differences are mechanism-consistent: the time-only provider supervises the exact observed bin inside the completion band and is slightly better where seconds-level supervision matters, which mirrors the anchor--restoration competition of Section~\ref{sec:provider}, while the typed provider keeps the only uniformly signed edge on OP-N@5, where its overplay support is supervised as one block.

\begin{table}[H]
\centering
\caption{Comparison of Typed and Time-Only Providers on the Main Tasks}
\label{tab:time-only-ablation}
\scriptsize
\setlength{\tabcolsep}{1pt}
\renewcommand{\arraystretch}{1.04}
\begin{tabular*}{\columnwidth}{@{\extracolsep{\fill}}lrrrr@{}}
\toprule
Provider & MAE & XAUC & Long-N@5 & OP-N@5 \\
\midrule
\multicolumn{5}{@{}l}{\emph{KuaiRec}} \\
Typed & \underline{3.5320$\pm$0.0019} & \underline{0.6817$\pm$0.0005} & \underline{0.8880$\pm$0.0006} & \textbf{0.8789$\pm$0.0007} \\
Time-only & \textbf{3.5297$\pm$0.0035} & \textbf{0.6823$\pm$0.0001} & \textbf{0.8886$\pm$0.0003} & \underline{0.8776$\pm$0.0008} \\
\midrule
\multicolumn{5}{@{}l}{\emph{KuaiRand-1K}} \\
Typed & \textbf{12.3583$\pm$0.0478} & \textbf{0.7361$\pm$0.0015} & \textbf{0.7180$\pm$0.0004} & \textbf{0.7421$\pm$0.0008} \\
Time-only & \underline{12.3654$\pm$0.0332} & \underline{0.7356$\pm$0.0009} & \underline{0.7175$\pm$0.0007} & \underline{0.7410$\pm$0.0013} \\
\midrule
\multicolumn{5}{@{}l}{\emph{WeChat21}} \\
Typed & \underline{13.0112$\pm$0.0043} & \underline{0.7011$\pm$0.0003} & \underline{0.8159$\pm$0.0008} & \textbf{0.8030$\pm$0.0006} \\
Time-only & \textbf{12.9808$\pm$0.0150} & \textbf{0.7012$\pm$0.0003} & \textbf{0.8164$\pm$0.0010} & \underline{0.8027$\pm$0.0009} \\
\bottomrule
\end{tabular*}
\par\smallskip
\begin{minipage}{\columnwidth}
\scriptsize
Both providers use matched encoders, output budgets, restoration, frozen summaries, raw-context inputs, and decision readouts. Each readout uses one million training impressions. Values are mean$\pm$sample standard deviation over seeds 2027--2029; bold and underline mark the better and worse mean within each dataset.
\end{minipage}
\end{table}
\FloatBarrier

\paragraph{Transfer.}
Under the frozen-transfer protocol of Section~\ref{sec:transfer-probe}, typed and time-only providers also perform similarly on the two transfer tasks, with all differences within or near seed-level variation.

\paragraph{Summary form.}
The third control fixes one frozen time-only provider and changes only the 27-dimensional summary passed to the readout.
The typed summary integrates the distribution over the four watch-ratio regions as above; the plain summary has matched dimension and uses no event boundary, consisting of nine log-time quantiles, nine duration-normalized quantiles, and nine global statistics of the time distribution such as its mean, spread, entropy, and peak mass.
The readout protocol matches the other controls.
Table~\ref{tab:summary-form} shows the two summaries perform on par: the plain summary is slightly better in MAE and XAUC on KuaiRec and slightly worse in MAE on WeChat21, while the typed summary keeps a small, uniformly signed edge on OP-N@5.
The typed discretization does not yield a consistent advantage over generic duration-relative statistics of the same distribution under these controls.

\begin{table}[H]
\centering
\caption{Comparison of Typed and Plain Summaries for a Frozen Time-Only Provider}
\label{tab:summary-form}
\scriptsize
\setlength{\tabcolsep}{1pt}
\renewcommand{\arraystretch}{1.04}
\begin{tabular*}{\columnwidth}{@{\extracolsep{\fill}}lrrrr@{}}
\toprule
Summary & MAE & XAUC & Long-N@5 & OP-N@5 \\
\midrule
\multicolumn{5}{@{}l}{\emph{KuaiRec}} \\
Typed & $3.5297{\pm}0.0035$ & $0.6823{\pm}0.0001$ & $0.8886{\pm}0.0003$ & $\mathbf{0.8776{\pm}0.0008}$ \\
Plain & $\mathbf{3.5066{\pm}0.0023}$ & $\mathbf{0.6831{\pm}0.0000}$ & $\mathbf{0.8898{\pm}0.0007}$ & $0.8770{\pm}0.0008$ \\
\midrule
\multicolumn{5}{@{}l}{\emph{KuaiRand-1K}} \\
Typed & $12.3654{\pm}0.0332$ & $0.7356{\pm}0.0009$ & $\mathbf{0.7175{\pm}0.0007}$ & $\mathbf{0.7410{\pm}0.0013}$ \\
Plain & $\mathbf{12.3646{\pm}0.0368}$ & $\mathbf{0.7361{\pm}0.0010}$ & $0.7174{\pm}0.0014$ & $0.7386{\pm}0.0022$ \\
\midrule
\multicolumn{5}{@{}l}{\emph{WeChat21}} \\
Typed & $\mathbf{12.9808{\pm}0.0150}$ & $\mathbf{0.7012{\pm}0.0003}$ & $0.8164{\pm}0.0010$ & $\mathbf{0.8027{\pm}0.0009}$ \\
Plain & $12.9952{\pm}0.0210$ & $0.7011{\pm}0.0002$ & $0.8164{\pm}0.0003$ & $0.8023{\pm}0.0007$ \\
\bottomrule
\end{tabular*}
\par\smallskip
\begin{minipage}{\columnwidth}
\scriptsize
Only the 27-dimensional summary changes. Values are mean$\pm$sample standard deviation over seeds 2027--2029; bold marks the better mean within each dataset.
\end{minipage}
\end{table}
\FloatBarrier

Replacing the provider supervision and replacing the summary form both leave performance essentially where it was, which is what the interface-level reading of Section~\ref{sec:matched-readout} predicts. The typed instantiation stays the default for the reasons given there.

\section{External Engagement Association}
\label{app:external-engagement}

Long relevance (completion or overplay) and overplay relevance are derived from the operational event types, not from independently recorded actions.
To assess their external association, we use the KuaiRand-1K test split's same-impression engagement mark $E$, defined as the union of separately logged like, follow, comment, and forward actions~\citep{gao2022kuairand}.
These current-impression actions are excluded from model inputs and do not supervise DSI's provider or ranking branch.
Table~\ref{tab:external-engagement} reports both aggregate association and duration-stratified overplay association.

\begin{table}[t]
\centering
\caption{Association Between Event-Type-Derived Relevance and Independently Logged Engagement on KuaiRand-1K}
\label{tab:external-engagement}
\scriptsize
\setlength{\tabcolsep}{1pt}
\renewcommand{\arraystretch}{1.05}
\begin{tabular*}{\columnwidth}{@{\extracolsep{\fill}}lrrrrr@{}}
\toprule
\multicolumn{6}{@{}l}{\textit{Panel A: Aggregate association}} \\
Relevance & Prev. & $\Pr(E\mid +)$ & $\Pr(E\mid -)$ & Lift & Corr. \\
\midrule
Long     & 19.41 & 3.79 & 1.13 & 3.35$\times$ & 0.0825 \\
Overplay & 10.27 & 4.27 & 1.35 & 3.17$\times$ & 0.0696 \\
\midrule
\multicolumn{6}{@{}l}{\textit{Panel B: Overplay association by duration quartile}} \\
Duration & $N$ & OP & $\Pr(E\mid\mathrm{OP})$ & $\Pr(E\mid\neg\mathrm{OP})$ & Lift \\
\midrule
Q1 ($d<12.17$s)          & 269,318 & 25.97 & 3.40 & 1.47 & 2.31$\times$ \\
Q2 ($12.17$--$35.63$s)   & 269,408 & 11.76 & 5.22 & 1.41 & 3.71$\times$ \\
Q3 ($35.63$--$104.77$s)  & 269,384 &  2.73 & 7.45 & 1.33 & 5.61$\times$ \\
Q4 ($d\ge104.77$s)      & 269,431 &  0.65 & 8.50 & 1.23 & 6.93$\times$ \\
\bottomrule
\end{tabular*}
\par\smallskip
\begin{minipage}{\columnwidth}
\scriptsize
$E=1$ denotes at least one like, follow, comment, or forward action. Prevalence and conditional-probability entries are percentages. In Panel~A, $+$ and $-$ denote the positive and negative class of the listed relevance definition, lift is $\Pr(E\mid+)/\Pr(E\mid-)$, and Corr. is the binary Pearson correlation. Duration stratification shows that the association remains positive within every quartile, while the sharp decline in overplay prevalence with duration confirms that overplay remains duration-dependent. The table supports criterion association only; it does not validate overplay as a replay label or an online business objective.
\end{minipage}
\end{table}

\section{WeChat21 Protocol and Feature Parity}
\label{app:wechat}

WeChat21 is the public short-video interaction log released by the WeChat Big Data Challenge 2021~\citep{wechat2021challenge}; it contains roughly 7.3M interactions between 20{,}000 users and 96{,}418 videos.
WeChat21 differs from KuaiRec and KuaiRand-1K in two dataset-level ways, both handled outside the core method and applied identically to every method.
First, the feed metadata has a pronounced pile-up at its 60-second duration cap: 20{,}054 of 106{,}444 feed records are exactly 60 seconds, while only 91 exceed it.
Because a capped duration is not an exact video endpoint, we evaluate the duration-below-cap subset, removing 1{,}267{,}015 impressions with $d\ge60$ seconds.
We also remove 13{,}075 extreme records with $y/d\ge10$.
The resulting dataset contains 6{,}037{,}792 impressions.
For reference, the corresponding full-duration data after the same extreme-ratio rule has early, mid, completion, and overplay frequencies of 39.08\%, 13.58\%, 11.63\%, and 35.71\%; on the duration-below-cap subset they are 36.88\%, 14.11\%, 12.60\%, and 36.41\%, respectively.
Both rules are fixed before model training and applied identically to every method.
DSI uses the same ordinal-integral value branch, ranking branch, and support-aware provider as on the other two datasets.

\paragraph{Shared model protocol.}
Within each dataset, every method uses the same processed split, raw-feature block, and evaluation code.
DSI contributes only the output of its own frozen estimator to its readout, just as each baseline contributes its native estimator output; this output is not an additional raw input supplied only to DSI.
The provider objective and the value and ranking branches follow the same design on all three datasets.

\paragraph{Feature parity.}
All methods read the same processed features through one loader: a shared numeric block and the same item-content fields.
Past-only user-history statistics (prior play time, completion, overplay, and engagement rates) are computed in user-log order and shifted by one impression, so current and later outcomes never enter the feature vector.
WeChat21 has no intra-day exposure timestamp, so we retain dataset order within each user-date pair.
Content-side tag and feed fields are native dataset attributes, not method components.

\end{document}